# Great Wall-like Water-based Switchable Frequency Selective Rasorber with Polarization Selectivity

Lingqi Kong, Xiangkun Kong, *Member, IEEE,* Shunliu Jiang, Yuanxin Lee, Lei Xing, and Borui Bian

*Abstract*—A water-based switchable frequency selective rasorber with polarization selectivity using the Great Wall structures is presented in this paper. The proposed structure comprises a container containing horizontal and vertical channels enabling dividable injection of water, and a cross-gap FSS. The novelty of the design lies in its switchability among four different operating states by injecting water or not into the water channels. When the container is empty, the structure acts as a polarization-intensive FSS with a -0.42 dB insertion loss passband at 3.75 GHz. When the horizontal channel is filled with water and there is no water in the vertical channel, this structure can be used as an FSR with single polarization selectivity. The FSR with -10 dB absorption band from 6.8 GHz to 18.8 GHz only allows certain polarized electromagnetic (EM) waves to pass at 3.1 GHz with an insertion loss of -0.78 dB, while another polarized EM wave cannot pass. When the container is full of water, the structure operates as an absorber with a reflection band below the absorption band, where neither of polarization EM waves can transmit. Besides, a reconfigurable water-based FSR automatic control system is built to achieve state switching and temperature constancy of the water within the container. Ultimately, a prototype of the presented design is fabricated, simulated and measured to verify the feasibility. This work has potential application in radome design to realize the out-of-band RCS reduction.

*Index Terms*— Frequency selective surface, rasorber, switchable rasorber, multifunctional, polarization selectivity

## I. INTRODUCTION

Frequency selective surface (FSS) has been widely applied in numerous military communication systems due to its spatial filtering characteristics [1]. The EM performance of FSSs usually depends on the constituting elements for the array. For example, FSSs using patches or apertures periodic units exhibit in-band reflection and transmission, respectively. Especially, bandpass FSS can transmit EM waves with low insertion loss in a specific frequency band but produces large reflection outside the passband [2]-[3]. To improve the stealth performance of antennas, the designs of bandpass FSS with reduced reflection from the reflection bands attract more and more attention. These new designs show absorptive frequency-selective transmission characteristics, and we academically call them frequency selective rasorber (FSR), where the "rasorber" is a combination of "radome" and "absorber" [4]. Under many military circumstances, it is desirable to employ a frequency selective rasorber for the sake of preventing being detected by a hostile reconnaissance system.

To date, existing FSRs can be separated into 2-D FSRs and 3-D FSRs categories based on their different structures. Most of the proposed 2-D FSRs are constructed from a lossless bandpass FSS with an absorptive surface that is usually loaded with electrical components or printed by high resistance ink [5]-[6]. In the previous studies, the vast majority of 2-D FSRs placed a transmission band and absorption band separately to ensure that the transmission performance wouldn't be affected by the lossy portion [5]-[9]. Several designs aiming to create a transmission band within an absorption band were proposed [10]-[13]. In these designs, the influence from the lossy components was minimized for the reason that losses were bypassed by using series LC resonators operating at the passband frequencies to short-circuit the resistors in the lossy layer. Since the concept of FSR was proposed [14], a considerable number of designs have been brought forward to improve the EM performance such as low-insertion loss [5] [13] [15]-[18], low profile [13] [19]-[20], wide absorption band [10] [21]-[22], etc. In order to obtain high-selective transmission performance and angular stability, 3-D FSRs consisting of a 2-D periodic array of multimode cavities were employed [23]-[26]. Compared with 2-D FSRs, although 3-D FSRs are even more complicated, their unit cell size can be shrunk to

Manuscript received November 10, 2020. This work was supported by the National Natural Science Foundation of China under Grant 62071227 and Grant 61601219, in part by the Natural Science Foundation of Jiangsu Province under Grant BK20160804, in part by the Open Research Program in China's State Key Laboratory of Millimeter Waves under Grant K202027, in part by the project of Key Laboratory of Radar Imaging and Microwave Photonics, Ministry of Education of China under Grant RIMP2020005, and in part by the Postgraduate Research & Practice Innovation Program of Jiangsu Province under Grant SJCX20_0070. *(Corresponding author: Xiangkun Kong.)*

L. Kong, X. Kong, S. Jiang and L. Xing are with the College of Electronic and Information Engineering, Nanjing University of Aeronautics and Astronautics, Nanjing, 211106, China (e-mail: 254815268@qq.com; xkkong@nuaa.edu.cn).
Y. Lee is with the College of Electronic Science, National University of Defense and Technology, Changsha, 410073, China (e-mail: 852648577@qq.com)
B. Bian is with the College of Information Science and Technology, Nanjing Forestry University, Nanjing, 210037, China (e-mail: borui_bian@sina.com)
X. Kong is also with the State Key Laboratory of Millimeter Waves, Southeast University, Nanjing, 210096, China

smaller than operating wavelength [27].

However, with the advent of the 5G era, active FSSs or FSRs represent a new challenge to traditional passive FSSs. To solve this disadvantage, active FSS has been extensively researched using several main methods, including electrical tuning, material tuning, optical tuning and mechanical tuning. There're several ways to achieve electrical tunability. PIN diodes and variable capacitors, as typical active elements, are widely employed in the implementation of FSSs with reconfigurability [28]-[30], tunability [31]-[33] and multifunctionality [34]-[37]. The same approach is applied to the design of reconfigurable or tunable FSRs as well [38]-[40]. Employing micro-electro-mechanical systems (MEMS) is also an effective way to realize active FSSs but to operate in higher frequency band [41]. For material tuning, liquid crystal and graphene are usually used to make FSSs tunable over a wide frequency range by changing the biased voltage or electrostatic field, respectively [42]-[43]. Ferrite-based magnetically tunable FSS and FSR in [44] and [45] have given material tuning another option. Recently, a different tact has been taken to realize active control with light assistance [46]. Unlike the three methods above, mechanical tuning is achieved simply and directly by changing the unit cells' geometric configuration, and a representative case is origami design [47]. Last but not least, liquid also has enormous potential in the area of switchable FSSs due to its excellent fluidity [48]. However, all the above designs have their defects, such as complex feeders and the narrow reconfigurable band for electrical tuning, complicated processing and high cost for material tuning, which are equally applied to optical tuning.

Water, as one of the most widely distributed resources on earth, has been used to design liquid-based antennas in the past few years [49]-[52]. It can also be used to design reconfigurable antennas due to its excellent fluidity, enabling the antenna to change its polarization or pattern by controlling the status of water [49]. More importantly, water has a high loss tangent (especially at high frequencies), which makes it a promising candidate for all-dielectric metamaterial absorber (MMA) [53]-[54]. Compared with the traditional absorbers, water-based absorbers have a lower profile, wider absorption band and lower cost. It makes optical transparent antenna and absorber possible [50]. Besides, water, as a dispersion medium, its permittivity varies with frequency and temperature. Such a unique feature makes it possible to design a thermally tunable water-based absorber. In practical engineering applications, water is easily available and eco-friendly. What's more, the specific heat capacity of water is relatively high, which is preferred for cooling or heating to achieve constant temperature.

Before our previous research [55], using water to design FSR had never been proposed. In [55], a water-based reconfigurable FSR was demonstrated for the first time. By taking advantage of water's fluidity, FSR can switch between Rasorber and FSS by injecting the water into the container or pumping it out. The dispersion characteristic of water was also utilized to realize the thermal tunability of FSR. By changing water's temperature in the container, the absorption band can be adjusted while remaining the passband steady. In this study, we make a further step on that. Inspired by the Great Wall structure, we ingeniously build two water channels in the horizontal and vertical arrangement at the same relative height. The two kinds of water channels are arranged in a woven structure that allows transverse and longitudinal channels to circulate water independently without affecting each other. The whole container is made by 3D printing technology. The novelty of the design lies in its reconfigurability among four different operating modes (dual-polarized FSS, x-polarized FSR, y-polarized FSR, and dual-polarized absorber) by injecting water or not in the transverse and longitudinal channels. So far, to the best of our knowledge, limited by complex circuit design and EM performance, polarization-reconfigurable or multifunctional FSR has not been given out yet. A reconfigurable water-based FSR automatic control system has also been given in this paper for state switching more efficiently. The establishment of intelligent automatic liquid control system has become an important breakthrough in this research work.

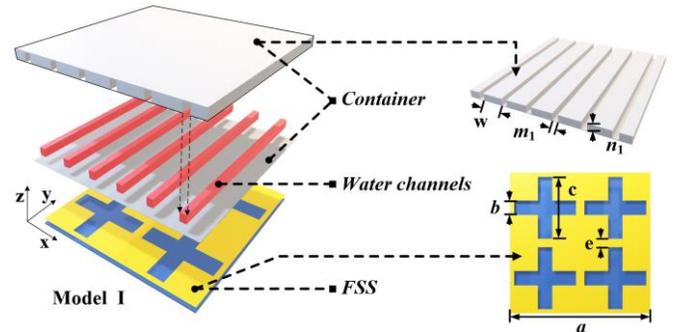

Fig. 1. The unit cell structure of the water-based FSR (model Ⅰ). (Structure dimension ($w = 9.03\ mm, m_1 = 2.3\ mm, n_1 = 3.1\ mm, a = 68\ mm, b = 30\ mm, c = 7mm, e = 4\ mm$.)

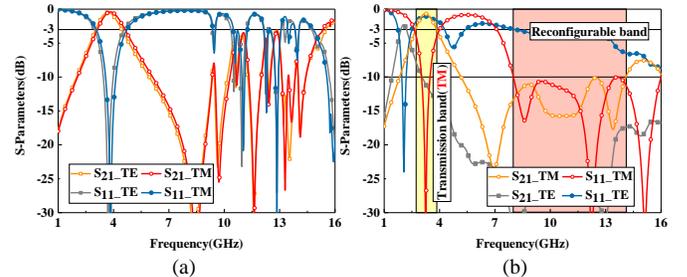

Fig. 2. The simulation results of the proposed FSR for two different working states: (a) "0" state and (b) "1" state.

## II. FSR DESIGN PROCEDURE AND CONTROL SYSTEM

### A. Switchable FSR Design Procedure

We start from a switchable water-based FSR (model Ⅰ) with single-polarization selectivity, displayed in Fig. 1, which consists of a cross gap FSS as the lossless-layer and several parallel rectangular water cylinders as the lossy-layer. Four cross slot structures of the FSS and six water channels compose one unit-cell of the FSR. FR$_4$ ($\varepsilon_r = 4.3, tan\delta = 0.025$) is chosen as the dielectric substrate with a thickness of $d_1$=1 mm for the FSS. These square-columnar water channels are filled in resin ($\varepsilon_r = 2.8, tan\delta = 0.0318$) containers made by 3D-printers. The S-parameters of the FSR (model Ⅰ) are simulated and the results are presented in Fig. 2.

As is well-known that, the imaginary part of the dielectric parameter of water increases with frequency, nearly 20 after 5GHz but leaving a relatively low value before 5 GHz. Such a particular dispersion characteristic makes water a perfect candidate for designing FSR with a transmission band above the absorption band. Furthermore, due to the fluidity of water, water-based FSR structure transforms significantly by changing the state (empty or full) of the container. Thus, as shown in Fig. 2 (a) and (b), the preliminary design (model Ⅰ) has two different working states "$i$" ($i$= 0,1), where "$i$" denotes the state of the container. When the container is empty, it works as a bandpass FSS with resonant frequencies at 3.68 GHz for TE-polarization and 3.83 GHz for TM-polarization. This slight difference comes from the structural asymmetry of the container. When the container is full of water, the designed FSR operates as a polarization-selective FSR. Due to the water channels' grating arrangement, the FSR transmits TM polarization waves and its -3 dB passband covers from 2.7 GHz to 3.8 GHz but reflecting TE polarization waves at the same frequency band. At the "1" state, a -10 dB refection band for TM polarization from 8 GHz to 14.1 GHz can be obtained. When it comes to the "0" state, the bandpass FSS reflects off-band EM waves and the ≥-3 dB reflection band is from 4.76 GHz to 14.8 GHz. Therefore, we define the intersection of the two frequency bands, which display distinguishing EM properties, as the reconfigurable band in Fig. 2(b).

An improved design (model Ⅱ) is given in Fig. 3 to achieve dual-polarization selectivity. Based on model Ⅰ, we add a layer of water channels that are perpendicular to the lower water channels. As shown in Fig. 4, four different states "$ij$" ($i$= 0, 1; $j$= 0, 1.), can be obtained with the injection or drawing of the water in the top and bottom channels, where "$i$" and "$j$" denote the states of upper and lower containers, respectively. As a result of the water channel's grating arrangement, "01" and "10" states show polarization selectivity. When the upper and lower channels are full of water, both TE and TM polarization EM waves are blocked at the "00" state's passband. However, as shown in Fig. 4 (b) and (c), the absorption band disappears. The addition of the upper structure not only broke the impedance matching condition of model Ⅰ but also weakens the water's absorption of EM waves. The main reason for the disappearance of the absorption band is that the water layer's absorption lies in the resonance between water and metal FSS, which requires the two to be set closely.

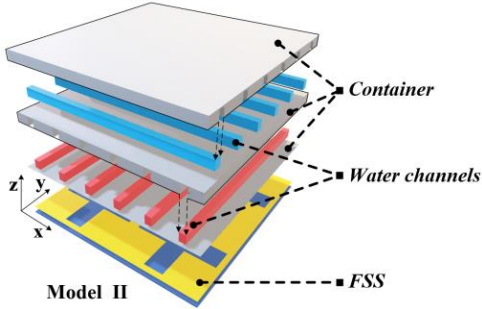

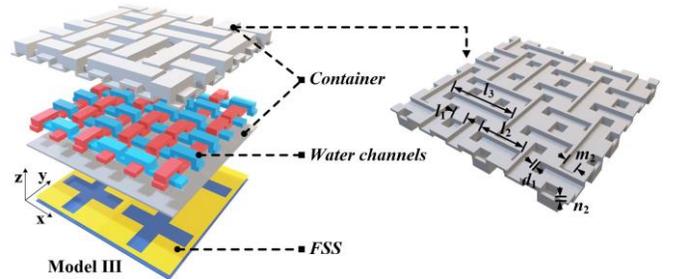

Fig. 5. The unit cell structure of the water-based FSR (model Ⅲ).

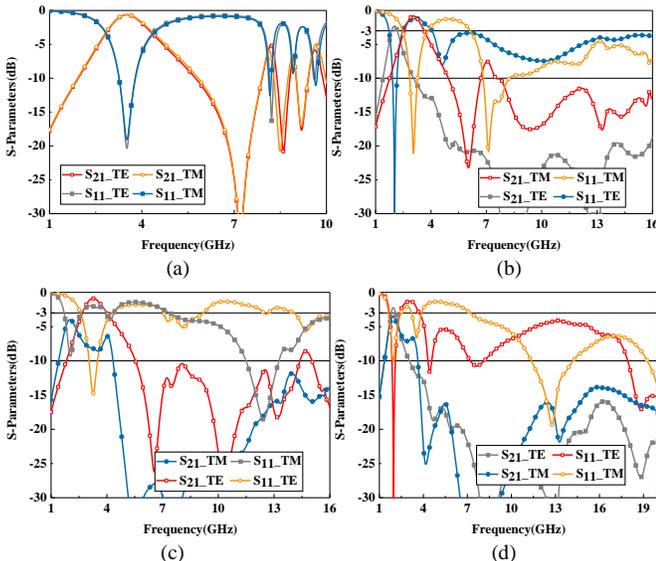

Fig. 3. The unit cell structure of the water-based FSR (model Ⅱ).

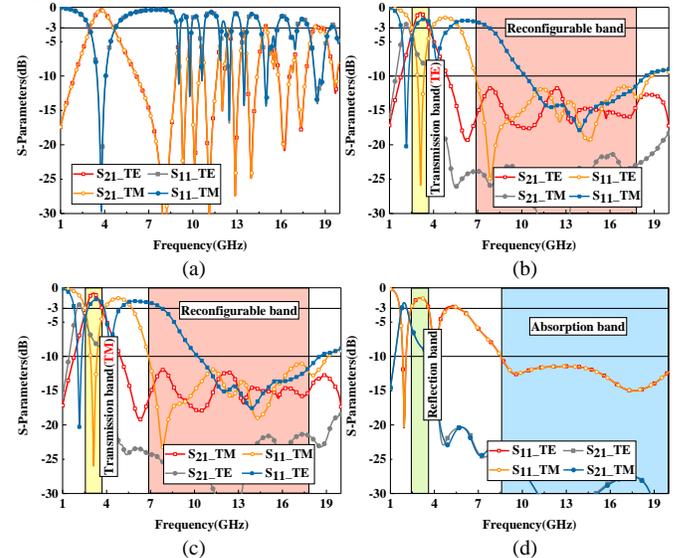

Fig. 6. The simulation results of the proposed FSR for four different working states: (b) "00" state (c) "01" state (d) "10" state (e) "11" state.

Thus, to improve the two working states' performance with polarization selectivity, inspired by the Great Wall structure, a final design (model Ⅲ) is given out in Fig. 5. The Great Wall-like structure is adopted in model Ⅲ to realize the horizontal (direction of *x-axis*) and vertical (direction of *y-axis*) water channels at the same relative height to the FSS. The final optimized dimensions are $l_1$= 5.6 mm, $l_2$= 16.37 mm, $l_3$= 27.13 mm, $m_2$= 4.3 mm and $n_2$= 1.8 mm. Considering the 3D printing technology requirement for the wall thickness, the wall thickness $d_1$ of the whole container is 1 mm, including the upper container for water passage and the lower cover plate.

Fig. 4. The simulation results of the proposed FSR for four different working states: (b) "00" state (c) "01" state (d) "10" state (e) "11" state.

As shown in Fig. 6, four independent operating states "$ij$" ($i$= 0, 1; $j$= 0, 1.) can be obtained by injecting water or not into the two kinds of channels, where "$i$" and "$j$" denote the states of transverse and longitudinal channels, respectively. As can be observed from Fig. 6(a), when no water is injected into the container, the proposed structure behaves as a simple polarization-intensive bandpass FSS with a -3 dB passband covering from 3 GHz to 4.6 GHz. It shows that the insertion loss is -0.42 dB for both TE and TM polarization at 3.75 GHz. Besides, the off-band EM waves are reflected. When the horizontal channels are filled with water and the vertical channels are empty, depicted in Fig. 6(b), the proposed structure works as an FSR with TE-polarization selectivity. It only allows the TE-polarization EM waves to pass but blocks the TM-polarization EM waves. At the "01" state for TE-polarization, the insertion loss is -0.78 dB at 3.1 GHz and the -3dB passband is from 2.6 GHz to 3.7 GHz. The bandwidth for refection less than -10 dB is 93.8% from 6.8 GHz to 18.8 GHz for TE-polarization. It is worth mentioning that, by injecting water into the container, there is a structural change in the proposed design, which causes the EM property to switch from reflection to absorption over a wide frequency band. Such reconfigurable band defined above is from 6.8 GHz to 17.8 GHz as shown in Fig. 6 (b) and (c). Because of the structure's precise symmetry under TE and TM polarization, the only distinction between the "01" state and the "10" state is the different polarization selectivity. When all channels are filled with water, it can be seen from Fig. 6(d) that the structure can be deemed as a wideband absorber. Under the "11" working state, both TE and TM polarization EM waves are reflected at the frequency band, which behaves as a passband at the "00" state.

TABLE I
WORKING STATES OF THE PROPOSED DESIGN

| Working State | Switchable T-R/T-A/R-A FULL-EMPTY state of the horizontal and vertical channels | | TE mode | TM mode |
|---|---|---|---|---|
| | Horizontal | Vertical | | |
| "00" | EMPTY | EMPTY | T -R | T-R |
| "01" | EMPTY | FULL | T -A | R-A |
| "10" | FULL | EMPTY | R -A | T-A |
| "11" | FULL | FULL | R -A | R-A |

T-R=transmission band below the reflection band;
T-A=transmission band below the absorption band;
R-A=reflection band below the absorption band.

In conclusion, the table Ⅰ has given out the detailed results of the four different working states.

*B. Reconfigurable Water-based FSR Control System Design*

The switching of various working states is realized through injecting water or not into the horizontal and vertical channels. However, in practical operation, manual injection greatly limits the efficiency of reconfigurability. More importantly, the

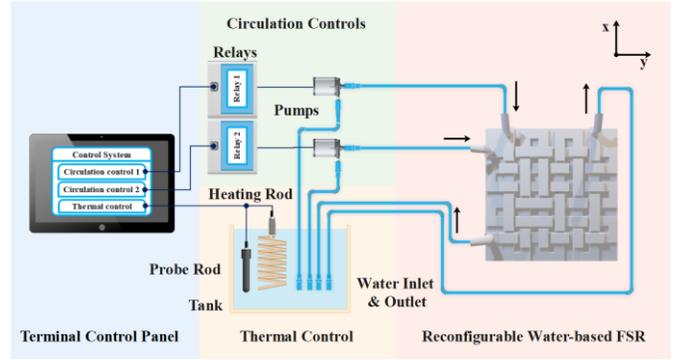

Fig. 7. The reconfigurable water-based FSR control system

temperature of the water used in the above established models is 25 ℃. According to the Debye formula [56]

$$\varepsilon(\omega,T) = \varepsilon_\infty(T) + \frac{\varepsilon_0(T) - \varepsilon_\infty(T)}{1 - i\omega\tau(T)} \quad (1)$$

where $\varepsilon_\infty(T)$, $\varepsilon_0(T)$ and $\tau(T)$ are the optical permittivity, static permittivity and rotational relaxation time, respectively. The real and imaginary parts of water vary with the temperature. If the temperature of the water inside the container is not 25 ℃, the impedance matching condition of the lossy layer at "01", "10" and "11" working states will be destroyed, leading to the deterioration of absorbing performance. Therefore, it is very important and necessary to build a temperature controller to keep the water temperature constant. Considering these imperious demands above, an automatic liquid control system has been built here to cooperate with water-based FSR so as to tune the circulation and temperature of water more efficiently, and also to monitor the water temperature constant at certain degrees. In previous studies, most mechanical tuning FSSs were preconfigured or tuned manually, limiting the tunning speed.

As depicted in Figure 7, the control system consists of three parts. Part 1 is a terminal control panel composed of a single chip microcomputer and a touch screen. Part 2 includes two independent water circulation controls comprising two water pumps, two relays and several water pipes. Part 3 is a water thermal control constituted of a water tank, a heating rod and a temperature probe rod. The water circulation controls are used to independently pump the water into the horizontal and vertical water channels within the FSR. When Circulation 1 is turned on, the pump connected to the Relay 1 starts to work, pumping the water out of the tank and injecting it into the vertical water channels. Circulation 2 works in a similar way. If we want to empty the horizontal or vertical channels, we can take out the corresponding pipes from the tank, and the air will be injected into these channels, replacing the water. The water thermal control adjusts the channel's temperature by heating the water in the tank to a specific value, thus realizing its thermal constancy. All these operations above are done on the touch screen, and the status of water-based FSR changes automatically with just a single finger slide. Of course, as the front end of human-computer interaction, the touch screen can also display real-time temperature changes and the difference value from the target temperature, convenient for operator monitoring.

## III. RESULTS ANALYSIS

*A. Distribution of field and power loss*

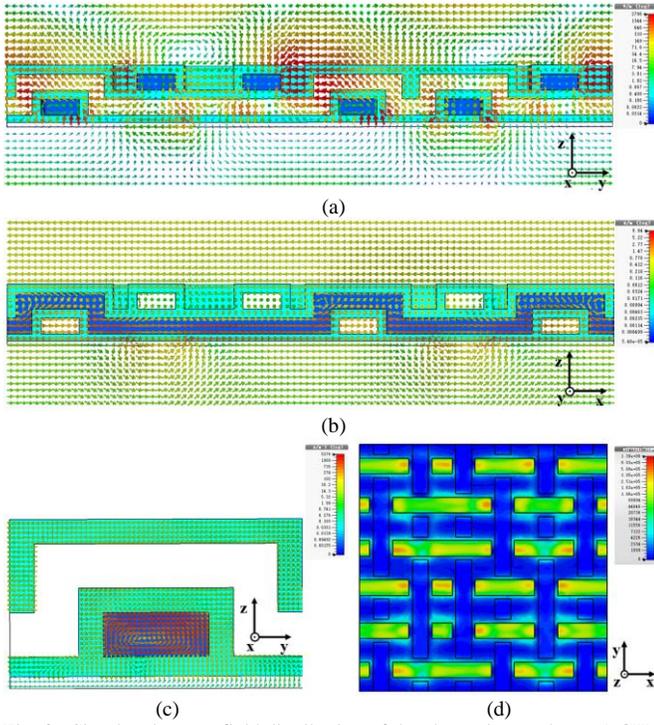

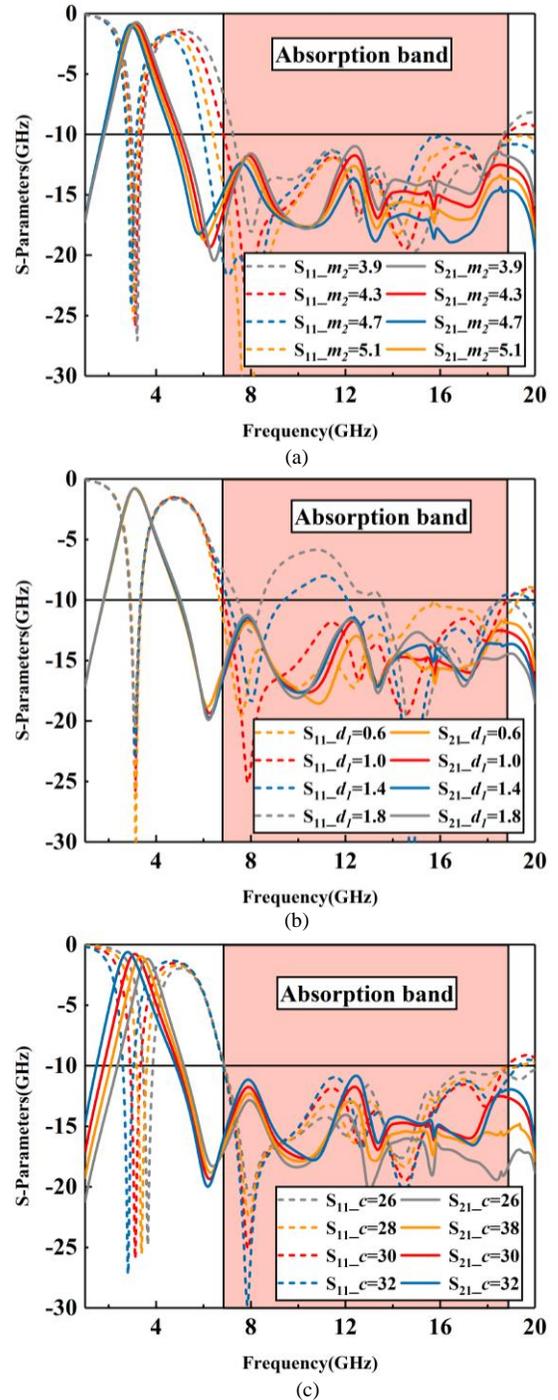

Fig. 8. Simulated vector field distribution of the absorption peak at 7.9 GHz. (a) E-field. (b) H-field. (c) current density. (d) the power-loss density.

In order to better understand the absorption mechanism of the designed FSR at the "01" state, the E-field and H-field distributions at 7.9 GHz resonance frequencies are simulated and carried out in Fig. 8 (a) and (b). It can be seen clearly that the E-field mainly concentrates around the water columns, as the permittivity's real part of the resin is extraordinarily smaller than that of the water. Simultaneously, the H-field mainly distributes in the bottom of the water and adjacently parallel to FSS. As depicted in Fig. 8(c), the water channel's cross section shows that the vortex-like displacement current loops are induced and formed by the reciprocal action between water and EM waves. These loops generate a strong magnetic field along the x-axis shown in Fig. 9(b). Because of the large loss tangent of water, most energy of the incident EM waves will be dissipated in the current loops localized in the water. Based on Fig. 8(a) and (b), intense electric and magnetic resonances generate from the coupling between the water channels and the incident fields at the absorption peak. Fig. 8(d) verifies the correctness of the above analysis. It can be seen that the energy loss is mainly concentrated in the water channels. As a result, water in the proposed FSR at the "01" state plays a significant role in absorbing EM waves.

*B. Parametric Study*

The EM performance of the designed water-based FSR is affected by many parameters. The decisive factors are the width of the water channels, the distance between the water layer and the FSS layer, and the relevant parameters of the FSS. In Fig. 9, the parameter sweeps of the proposed FSR are given to better explain the influence of parameters on the absorption and transmission performance. As shown in Fig. 9(a), the width $m_2$ of the cross-section of the water channels is the main parameter affecting the absorption bandwidth. Therefore, the absorption bandwidth can be expanded by increasing the $m_2$, but not be unlimitedly widened as it also affects the insertion loss. It can be seen from Fig. 9(b) that the distance (i.e., the thickness $d_1$ of the lower container) between the water layer and the FSS layer is the main factor affecting the absorption performance. The shorter the distance, the stronger resonant it will generate. Unfortunately, limited by the minimum thickness requirements of 3D-printing technology, the value of $d_1$ can only be minimized to 1 mm. As Fig. 9(c) depicted, the length $c$ of the cross-gap of the FSS not only affects the transmission frequency point and the insertion loss, but also affects the absorption bandwidth to a certain extent. Although it seems that the bandwidth can be expanded and the insertion loss reduced by increasing the length $c$ to achieve a double harvest, the increase of $c$ directly affects the absorption performance of the "11" working state, making the "11" state change from R-A to R. The parameter sweeps can indirectly verify the correctness of the aforementioned analysis. In general, the water layer is the main factor affecting the absorption of EM waves' energy, but the coupling between the water layer and the FSS layer also does the same job.

Fig.9. The effect of different parameters on the FSR performance at "01" working state under TE-polarization. (a) Width $m_2$ of the cross section for the water channels. (b) The thickness $d_1$ of the lower container. (c) The length $c$ of the cross-gap of the FSS.

## IV. EXPERIMENTAL RESULTS

As depicted in Fig. 10(a), the reconfigurable water-based FSR control system has already been built. The STM32H743 is used as the core-board, with a high-performance 4.3-inch capacitive touch screen module (ALIENTEK-4.3 TFTLCD) shown in Fig. 10(b). The temperature of water in the tank is logged by using the DS18B20 temperature sensor.

The proposed rasorber, with an overall dimension of 380 mm × 380 mm, has been fabricated and measured to verify the simulations. The detailed photo of the FSR has been given in

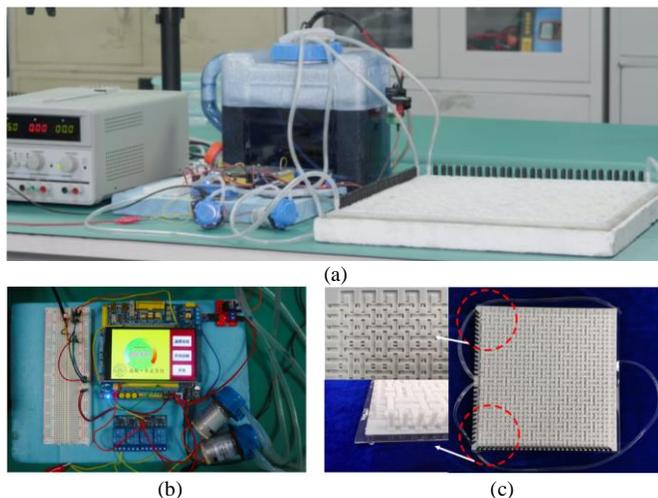

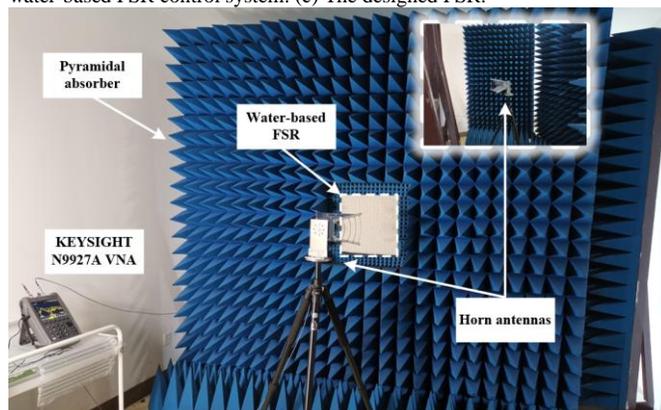

Fig. 10. (a) Complete view of the whole prototypes. (b) The reconfigurable water-based FSR control system. (c) The designed FSR.

Fig. 11. Measurement setup.

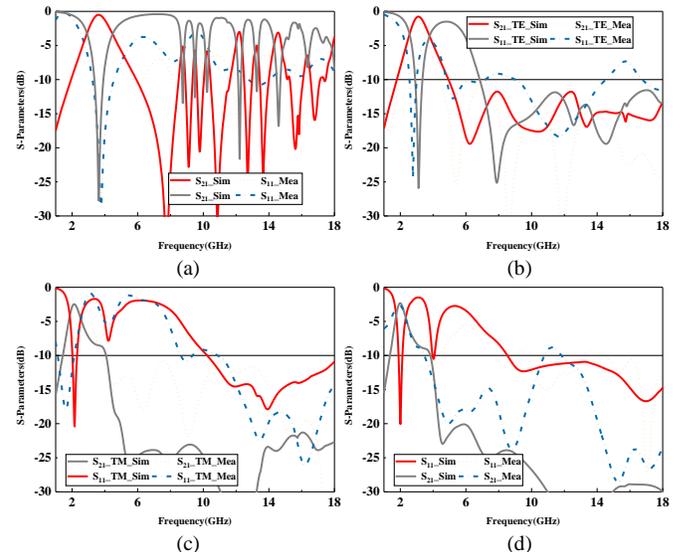

Fig. 12. Measured and simulated results at different working states. (a) "00" state. (b) "01" state under TE-polarization. (c) "01" state under TM-polarization. (d) "11" state.

Fig. 10(c). Considering the need to encapsulate water and avoid water leakage, the actual size of container is a little larger than the proposed design, thus leaving room for the glass glue. The whole container is 3-D printed by photosensitive resin, and all inlets and outlets are arranged around the structure.

To perform the measurement, the free-space technique is adopted by using the pyramidal absorber, and the measurement setup is shown in Fig. 11. The experimental results of the water-based FSR at "00", "01" and "11" working states are compared with the simulated ones in Fig. 12. Since the simulation results of "01" and "10" states are identical except for the opposite polarization selectivity, this situation is also applicable to the real test. So "10" state is not shown here to reduce the complexity of the image. It appears that the simulated and measured results show great agreement, especially for the main transmission frequency point. However, in practice, a large amount of glass glue was used for water encapsulation, which was not considered in the simulation. The addition of glue causes the impedance mismatch at a certain reflection frequency band and results in the absorbing frequency point deviation. It's also important to note that there will be few inevitable bubbles in the container, which leads to the test error. In general, the proposed water-based switchable FSR with polarization selectivity meets the requirements of EM performance.

## V. CONCLUSION

In this paper, a Great wall-like water-based switchable FSR with polarization-selectivity and its automatic control system were presented. The proposed water-based FSR possesses multifunctionality, including four different and independent working states controlled by injecting water or not into the horizontal and vertical channels. The whole water injection is completed by an intelligent automatic liquid control system equipped with a thermostat. It's a breakthrough to achieve interdisciplinary by endowing the automatic system's efficiency to FSR. So far as we know, no FSR with polarization-selectivity has been presented. The simulation results at "01" and "10" states show that the water-based FSR has a broader absorption band than traditional FSRs. The switching between operating states through a sizeable structural change within a unit brings a wider reconfigurable band from 6.8 GHz to 17.8 GHz. Meanwhile, the switching speed can be significantly improved by working with an automatic control system. This work successfully demonstrated that water-based FSR with polarization selectivity could be used in a smart stealth system, which meets the increasingly complex EM environment requirements.